\documentclass[aps,showpacs,amsmath,amssymb,12pt]{revtex4}
\usepackage{latexsym}
\usepackage{bm}

\def\a{\alpha}

\def\k{\kappa}

\def\be{\begin{equation}}
\def\ee{\end{equation}}

\begin{document}

\title{Conformal Properties of Charges in Scalar-Tensor Gravities }

\author{S. Deser}
\email{deser@brandeis.edu}
\affiliation{Department of Physics, Brandeis University, Waltham, MA 
02454, USA \\
and \\
Lauritsen Lab, Caltech, Pasadena CA 91125, USA }

\author{Bayram Tekin}
\email{btekin@metu.edu.tr}
\affiliation{Department of Physics, Faculty of Arts and Sciences,\\
             Middle East Technical University, 06531, Ankara, Turkey}

\date{\today}

\begin{abstract}
We study the behavior under conformal transformations of energy and other 
charges in generic scalar-tensor models. This enables us to conclude that 
the ADM/AD masses are invariant under field redefinitions mixing metric 
and scalar despite the permitted slow asymptotic falloff of massless 
scalars.
\end{abstract}

\maketitle

\section{\label{intro} Introduction}

Gravitational models involving massless scalars and (possibly) a 
cosmological constant can cause complications in applying otherwise 
well-understood energy and other definitions of physical quantities. In 
part, this is due to the scalars' slow allowed asymptotic falloff, 
generically as slow as that of the metric, $1/r^{D-3}$, along with the 
possibility of scalar-tensor field redefinitions, 
$g_{\mu \nu} \rightarrow g_{\mu \nu} f(\phi)$. One well-know example of 
deviation from pure Einstein behavior is the violation of the equivalence 
principle: the coefficients of the leading asymptotic terms of $g_{00}$ 
and $g_{i j}$ become unequal; the former define the Newtonian force, the 
latter the system's energy. Here, the potential difficulty is that, while 
energy for asymptotically flat (ADM \cite{adm} ) or asymptotically (A)dS 
(\cite{ad,dt1,dt2}) spaces is defined as though scalars are just another 
form of matter, their formal role can be shifted by the above field 
redefinitions that do not alter the physics 
any more than any other such field redefinitions in field theories. We 
show 
here that indeed, the correct physical quantities are invariant as 
desired, if not transparently so.

\section{\label{action} The generic action}

The most general second-order scalar-tensor theory in $D$ dimensions 
involves four arbitrary functions of the scalar field,
\begin{eqnarray}
S= \frac{1}{2\kappa}\int d^D x{\sqrt{-g}}\,U(\phi)\Big \{ R( g) + 
2 \Lambda_0 - 
W(\phi) \partial_\mu \phi \partial^\mu \phi - V(\phi) + 
H(\phi){{\cal L}}_{\mbox{m}} \Big \}, \label{stringframe} 
\end{eqnarray}
where  ${\cal{L}}_{\mbox{m}}$ represents all the matter besides the 
scalar. One-standard-field redefinition to the Einstein frame removes the 
overall function $U(\phi)$   
\be
g^E_{\mu \nu} \equiv U(\phi)^{\frac{2}{D}}g_{\mu \nu}. 
\label{scaling}
\ee
leading to 
\be
S= \frac{1}{2\kappa}\int d^D x{\sqrt{-g^E}}\Big \{ R(g^E)  + 
2 \Lambda_0  \Big \} + S_M,
\label{einsteinframe}
\ee
where $S_M$ now has the form
\be
S_M =\frac{1}{2\kappa}\int d^D x{\sqrt{-g^E }}\Big \{ A(\phi)\partial_\mu 
\phi 
\partial^\mu \phi + X(\phi) + 
Z(\phi){\cal{L}}_{\mbox{m}} ] \Big \}.
\ee
Obviously, a rescaling of $\phi$ reduces its kinetic term to free form, 
leaving just two arbitrary functions, the scalar self-interaction and  a 
possible coefficient of the matter action. We emphasize that this 
describes the same physics as (\ref{stringframe}) 
via a different metric variable. But the conserved and diffeo-invariant 
charges of the theory, similarly defined in terms of the metric variables 
each frame, do not obviously have the same value in each ( for the same 
physical configuration, of course), so one must verify this 
fact. [One can also study other conformal properties of the spacetimes. 
For example in \cite{jk} it was shown that the surface gravity and the
temperature of a stationary black hole are invariant under conformal   
transformations that approach  unity at infinity. ]

We have assumed here that the cosmological term is not altered by $V$, 
{\it {i.e}}, that $V(0) = 0$, something that we may always arrange or else 
just 
redefine things beforehand. In that case, the UV term would not change the 
value $\Lambda_0$ either, as long as $U(\phi)$ does not rise, $U(0)= 
const$.~\footnote{ A second issue is one mentioned in \cite{hmtz,clp}, 
namely 
possible 
scalar configurations whose kinetic and potential terms make divergent 
contributions to the energy. This would require a more careful choice of 
background with respect to which the energy is to be measured.} 

\section{\label{charges} Conserved Charges }

The conserved charges for the action (\ref{einsteinframe}) were derived in 
\cite{ad,dt1,dt2} for arbitrary (including vanishing) $\Lambda_0$. The 
result, for any of them (depending on the particular Killing vector) is 
\begin{eqnarray} 
Q^{\mu}(\bar{g^E}, \bar{\xi^E}) = &&\frac{1}{4 \,
\Omega_{D-2} \, G_{D}}\int_{\partial M}dS_i\sqrt{\bar{g^E}}
\Big \{ \bar{\xi^E}_\nu
\bar{\nabla}^{\mu}h_E^{i \nu} -\bar{\xi^E}_\nu 
\bar{\nabla}^{i} h_E^{\mu\nu}
+\bar{\xi_E}^\mu \bar{\nabla}^i h^E -\bar{\xi_E}^i 
\bar{\nabla}^\mu h^E   \nonumber \\ 
& + & h_E^{\mu \nu}\bar{\nabla}^i \bar{\xi^E}_\nu - h_E^{i \nu}\bar{\nabla}^\mu
\bar{\xi^E}_\nu + \bar{\xi_E}^i 
\bar{\nabla}_{\nu}h_E^{\mu \nu} 
-\bar{\xi_E}^\mu
\bar{\nabla}_{\nu}h_E^{i \nu} + h^E\bar{\nabla}^\mu 
\bar{\xi_E}^i \Big \}
\label{charge} 
\end{eqnarray} 
where we have split 
$g^E_{\mu \nu} \equiv \bar{g^E}_{\mu \nu} + h^E_{\mu 
\nu}$ and defined   $ h^E =h^E_{\mu \nu} 
\bar{g_E}^{\mu \nu}$; 
$\bar{\xi^E}$ is a Killing vector with respect to the asymptotic 
(A)dS or flat background $\bar{g}_E$. The 
integral 
is to be 
computed on a spatial hypersurface at infinity (only then is the 
expression diffeomorphism invariant). The covariant 
derivatives are also with respect to the Einstein-frame metric. 
For flat background, (\ref{charge}) reduces to the ADM mass \cite{adm}, 
but in arbitrary, rather than the Cartesian coordinates. If the 
background is dS, then as is well-understood, the cosmological horizon 
forbids timelike Killing vectors  
outside, and one can only deal with systems localized within the horizon.
  
To return to our energy problem, consider the inverse transformation to 
(\ref{scaling}), from Einstein to generic frame. This leads to the 
following scaling of the integrand of the conserved charge (\ref{charge}) 
\begin{eqnarray}
\sqrt{g} q^{i\mu}(\xi) &=& 
U^{-\frac{2}{D}} \sqrt{\bar{g}^E} \Big \{ q^{i \mu}(\xi^E)-
\frac{3}{D} \xi^E_\nu h_E^{i\nu}\partial^\mu \log U +
\frac{3}{D}\xi^E_\nu h_E^{\mu \nu}\partial^i \log U 
\nonumber \\
&&-\frac{D-1}{D}\xi_E^i h_E^{\mu \nu}\partial_\nu \log U + 
\frac{D-1}{D}\xi_E^\mu h_E^{i\nu} \partial_\nu \log U  
\Big \},
\label{confcharge}
\end{eqnarray}
and immediately proves our desired result: if $U(\infty)= 1$ then  
$g_{\mu \nu}$ and  $g^E_{\mu \nu}$ have the  `same` charges.  If on the other 
hand $ U(\infty)$ is some arbitrary constant, then the charges of these two 
metrics differ by a multiplicative constant. This result means that all 
charges given in the form (\ref{charge}) are invariant.

Our construction is quite generic: Higher curvature models, to which we 
now turn , can also be handled in a similar fashion as above. In 
\cite{dt1,dt2}, we constructed conserved charges 
in generic higher curvature gravity models. First let us consider 
the theories known as $F(R)$ gravities recently suggested in connection 
with the accelerated expansion of the Universe.
Their actions read 
\be
S=\frac{1}{2\k^2}\int d^4x\sqrt{-g}\left (F(R)+ 2 \Lambda_0 \right ) 
+S_{m},
\label{highercurv}
\ee
where $F(R)$ is a function of the Ricci scalar $R$ {\it only }. For 
example it could 
be $R + \mu^4/ R$. 

The model (\ref{highercurv}) can be re-expressed 
as a scalar-tensor theory
\be
S=\frac{1}{2\k^2}\int d^4x\sqrt{-g}\left (F(\phi)+F'(\phi)(R-\phi) 
+2\Lambda_0 \right )+
S_{matter},
\label{highercurv2}
\ee
where $F'(\phi)=dF/d\phi$ and  $F''(\phi)\neq 0$. Furthermore, assuming 
that 
$F'(\phi)g_{\mu\nu}=g_{\mu\nu}^E$ 
the action can be reduced to that of the
scalar field minimally coupled to the Einstein 
gravity. The explicit form of the action is not needed here, as we are 
only interested in the fact that the conserved charges once again are 
given as (\ref{charge} ). 

Lastly, let us consider quadratic models of the form
\be
I = \int d^D\, x \sqrt{-g} \Big \{ \frac{R}{\kappa} +
\a R^2 + \beta R^2_{\mu\nu} +
\gamma (R_{\mu\nu\rho\sigma}^2 -4R^2_{\mu\nu}+R^2 ) \Big \}.
\label{quadraticaction}
\ee
Rescaling of the metric ({\ref{scaling}}) maps the above model to a highly 
complicated one, whose action we do not need here. But, in the line of 
the discussion above, let us see how the charges transform under such a 
scaling. In \cite{dt1,dt2}, it was shown that 
the for asymptotically AdS 
spaces, the non-trivial part of the charge is given as 

\be
Q^\mu  =  \Big \{
\frac{1}{\kappa } + \frac{4\Lambda D\a}{D-2} + 
\frac{4\Lambda\beta}{D-1} +
\frac{4\Lambda \gamma (D-4)(D-3)}{(D-2)(D-1)}\Big \}
 \times 
Q^\mu_{Einstein}, 
\ee
where $Q^\mu_{Einstein}$ is given by (\ref {charge}). Therefore, the 
conserved charges of the theory (\ref{quadraticaction}) transform as 
in (\ref{confcharge}) under conformal scalings of the metric.
Any higher curvature model, including actions that depend on inverse powers 
of the scalar invariants constructed from the Ricci and Riemann tensors,  
can be handled this way. But of course getting the surface form of the 
energy expressions will be tricky in some cases.

\section{\label{summary} Summary }

We have studied the properties of 
conserved charges in various gravity models, such as Einstein, higher 
curvature models and 
scalar-tensor theories, both for asymptotically AdS and flat 
spacetime: They  are conformally invariant  as long as the conformal 
factor goes to unity at infinity. Our formalism also relates the charges of 
a generic scalar-tensor theory to Einstein's theory minimally coupled to scalar 
fields. Amongst other open problems, we hope to return to the treatment of 
solutions with scalar fields that radically alter the asymptotics.

\section{Acknowledgments}
The work of S.D.\ is supported by NSF grant PHY 04-01667; that of
B.T.\ by the ``Young Investigator Fellowship" of Turkish Academy of
Sciences (TUBA) and by the TUBITAK Kariyer Grant no 104T177.

\end{document}